\documentclass[a4paper,12pt,reqno,superscriptaddress,nofootinbib]{revtex4}

\usepackage[centertags]{amsmath}
\usepackage{amsfonts}
\usepackage{amssymb}
\usepackage{amsthm}
\usepackage{newlfont}
\usepackage{stmaryrd}
\usepackage{mathrsfs}
\usepackage{mathtools}
\usepackage{euscript}
\usepackage{graphicx}
\usepackage{color}
\usepackage{soul} 
\usepackage[font=small,labelfont=bf,justification=RaggedRight,singlelinecheck=false]{caption}
\usepackage{hyperref}
\usepackage{physics}
\usepackage{siunitx}
\usepackage{textcomp}
\usepackage{grffile}

\newcommand{\bbR}{{\mathbb R}}

\newcommand{\bbZ}{{\mathbb Z}}

\newcommand{\opunit}{\text{1}\kern-0.22em\text{l}}

\DeclareMathAlphabet{\mathpzc}{OT1}{pzc}{m}{it}

\def\mbf{\mathbf }

\let\oldsqrt\sqrt
\def\sqrt{\mathpalette\DHLhksqrt}
\def\DHLhksqrt#1#2{%
	\setbox0=\hbox{$#1\oldsqrt{#2\,}$}\dimen0=\ht0
	\advance\dimen0-0.2\ht0
	\setbox2=\hbox{\vrule height\ht0 depth -\dimen0}%
	{\box0\lower0.4pt\box2}}

\begin{document}
	\title{Photon frequency diffusion process}	
	\author{Guilherme Eduardo Freire Oliveira, Christian Maes and Kasper Meerts\\ {\it Instituut voor Theoretische Fysica, KU Leuven}}

\begin{abstract}
	We introduce a  stochastic multi-photon dynamics on reciprocal space.  
	Assuming isotropy, we derive the diffusion limit for a tagged photon to be a nonlinear 
	Markov process on frequency.  The nonlinearity stems from the stimulated 
	emission.  In the case of Compton scattering with thermal electrons, the 
	limiting process describes the dynamical fluctuations around the Kompaneets 
	equation.  More generally, we construct a photon frequency diffusion 
	process which enables to include nonequilibrium effects.  Modifications of 
	the Planck Law may thus be explored, where we focus on the low-frequency 
	regime.
\end{abstract}
\maketitle

\section{Introduction}
Time-dependent and nonequilibrium dynamics of photons are of increasing interest in a wide range of subjects. Quantum electromagnetic fields can be manipulated  to produce photonic lattices (spatial network of coupled photon modes) where time-dependent driving leads to a breaking of time-reversal invariance (synthetic magnetic field) \cite{koch,fang,roushan}.  Same thing for wave guides where the input field may be driven by either a laser or microwave generator, imposing a nonequilibrium boundary condition on the propagating photons \cite{pozar,roy}.  In quantum optomechanical studies photon gases may be enclosed in cavities with time-dependent geometry such as from vibrating walls, \cite{metaphotonics}.    In another domain and since much longer, plasma physics has dealt with the problem of understanding the origin of suprathermal tails \cite{supra1,banerjee} and high-energy cosmic radiation as a result of the scattering of electrons with turbulent electromagnetic fields \cite{fermi, sturrock,brin}.  Finally, in Early Universe cosmology, the dynamical origin of the main features of the cosmic background radiation (CMB) remains of central importance, especially to understand (possible) deviations from homogeneity \cite{dark} or from the Planck Law \cite{arcade1,arcade2,edges,arca}.\\
As understood from the pioneering days of statistical mechanics and in analogy with Brownian motion, we may expect that the studies mentioned above benefit from modeling fluctuations, {\it i.e.}, to identify the random motion in photon frequency space. For example, considering a mesoscopic level of description, we become able to insert nonequilibrium effects on the single-photon level allowing to discover meaningful modifications of the Planck Law.\\

In the present paper  we derive a (stochastic) diffusion process for a tagged photon, that can be seen as the diffusion limit of a multi-photon hopping dynamics in reciprocal space.  Stimulated emission results in the nonlinearity of the Markov diffusion process.\\ 
  We first apply our construction to build a Kompaneets process, {\it i.e.}, the stochastic (single-photon) dynamics that has the well-known Kompaneets equation as its nonlinear Fokker-Planck equation.
That new fluctuation dynamics  is the counterpart to the analysis of Kompaneets in \cite{kompa} and makes our first main result: the derivation of the frequency diffusion process for a tagged photon in a plasma dominated by Compton scattering.  Similarly, many other processes such as double Compton and  \textit{Bremsstrahlung} can be considered, and thanks to our setup, are easily added to the dynamics, turning it into a fully nonlinear reaction-diffision dynamics.  Not only does that process complement (in some precise sense) the Kompaneets equation, it allows also to see beyond, leading to our second main contribution.  A possible avenue indeed that gets opened, is the implementation and exploration of nonequilibrium effects modifying the Planck law, via physically motivated interventions on the single-photon level. Our work explores how changing drift and adding diffusion may change the low-frequency distribution, where the relaxation times in the Kompaneets process are largest, hence most vulnerable to nonequilibrium amendments.\\
As a final remark, it should be noted in all these cases that we start from a multi-photon diffusion process in frequency space.  It means to include the stimulated emission in the reactivities. That presents a nontrivial aspect in both the theoretical and computational analysis, leading finally to important nonlinearities in the corresponding Fokker-Planck equation. However, our results indicate that the processes simulate indeed the Kompaneets equation and its extensions, in the sense that the empirical density of photons follows these equations.\\

In the next section we describe the Kompaneets equation for relaxation to the Planck law via Compton scattering.  In Section \ref{ksp} we introduce the Kompaneets {\it process}, describing the hopping of photons in reciprocal space.  Its diffusion limit for a tagged photon gives a nonlinear Markov process, described in Eq. \eqref{kp-ito2}.   That simulation is explained in Section \ref{simul}, where the results are shown in terms of the time-dependence of the spectral density. We confirm the validity of the simulation scheme by verifying relaxation to the Planck law along the Kompaneets equation. There, we also show the appearance of a condensate when the number of photons is taken to be large enough, an effect already observed in \cite{levermore}. Other non photon-number preserving radiation processes are added in Section 
\ref{rbdc}. Bremsstrahlung and double Compton scattering are described there as low-frequency corrections. These are reactive mechanisms to control photon-number that can be handled and taken effectively in the simulation.   Section \ref{extk-sec} discusses more general photon-number processes, extending the Kompaneets process to implement two types of low-frequency modifications, either in the drift or in the diffusion. They are nonequilibrium features in the photon dynamics, effectively implementable on the single-photon level that we simulate. Finally, in Section \ref{simex}, we explain the simulation details for that extension (including reactive mechanisms). That illustrates in great detail how nonequilibrium features change the stationary solution to enhance low-frequency occupation (yielding there a higher effective temperature).

\section{Kompaneets equation}

The Compton effect is a quantum process in which photons scatter from free electrons.  It eventually leads to the relaxation of the photon distribution to that of the Planck radiation law.
The Kompaneets equation 
\begin{equation}\label{ke}
	\omega^2\frac{\partial n}{\partial t}(t,\omega)= \frac{n_e\sigma_T 
		c}{m_e c^2}\frac{\partial }{\partial \omega}\omega^4\left\{k_B T 
	\frac{\partial n}{\partial \omega}(t,\omega) + 
	\hbar\left[1+n(t,\omega)\right]n(t,\omega)\right\}
\end{equation}
 describes that relaxation towards equilibrium of a photon gas in contact with a nondegenerate, nonrelativistic electron bath in thermal equilibrium at temperature $T$. Here, $n(t,\omega)$ is the average occupation number  at frequency $\omega$ of the photon gas at time $t$; stationarity is then achieved when $n(t,\omega)$ reaches the Bose-Einstein distribution. Apart from the usual constants in \eqref{ke}, we recognize $\sigma_T$ as the Thomson total cross section and $n_e,m_e$ as the density and mass of the electrons, respectively.  The induced Compton scattering \cite{liedahl, blandford}  leads to the nonlinearity (in the second term) appearing in the Kompaneets equation \eqref{ke}.\\
 In 1957 Kompaneets \cite{kompa} gave a mesoscopic derivation of \eqref{ke}, starting from a semi-classical Boltzmann equation. It remains essential for the understanding of the CMB spectrum and related phenomena such as the Sunyaev-Zeldovich effect \cite{sunyaeveffect,sunyaev}.  The understanding of the Kompaneets equation has been evolving over the years and excellent reviews include \cite{practical,gui,zeldovich}. We emphasize that many extensions to the equation exist.  For example, the isotropy condition for the distribution  has been relaxed in \cite{buet, pitrou}, while nonrelativistic extensions can be found in \cite{barbosa, brown, itoh, itoh2, cooper, kohyama1, kohyama2, kohyama3}.  More recently in \cite{paper} we addressed some consistency problems in Kompaneets' original framework \cite{kompa}.\\

For simplicity it is convenient to use the dimensionless counterpart of \eqref{ke} in order to introduce (in the next Section) the Kompaneets {\it process}, which describes the hopping of photons in reciprocal space.\\ 
The dynamics of the average photon occupation number $n(t, x)$ at dimensionless frequency $x= \hbar \omega/k_B T$ in a thermal environment with temperature $T$, is obtained from equation \eqref{ke} as
\begin{equation}\label{ake}
x^2\frac{\partial n}{\partial t}(y,x) = \frac{\partial }{\partial x}x^4\left\{
\frac{\partial n}{\partial x}(y,x) + 
\left[1+n(y,x)\right]n(y,x)\right\}
\end{equation}
where we have defined the dimensionless Compton optical depth
\[y = \frac{ k_B T }{m_e c^2} n_e \sigma_T c \, t =  \frac{ t}{\tau_C}.\]
Here, $\tau_C$ is the  characteristic time in which photons change their frequency due to Compton scattering with thermal electrons,
\begin{equation}\label{shift}
\left\langle\frac{1}{2\tau}\left(\frac{\Delta\omega}{\omega}\right)^2\right\rangle\approx \frac{ k_B T }{m_e c^2}\; \frac 1{\tau}= \frac{1}{\tau_C}
\end{equation}
where $\tau = l/c$ is the average collision rate, related 
to the mean free path of photons $l=(n_e\sigma_T)^{-1}$. For simplicity, we will take our units such that $\tau_C = 1$, identifing $y$ with $t$.\\
Note that the
\begin{equation}\label{cc}
n_\mu(x) = \frac 1{e^{x-\mu}-1}, \qquad \mu\leq 0
\end{equation}
are stationary solutions of \eqref{ake}: $\frac{\dd n_\mu}{\dd x}(x) + [ 
1+n_\mu(x) ] \,n_\mu(x)=0$.\\

Central for our purposes to simulate the Compton scattering on a mesoscopic level is to write the Kompaneets equation \eqref{ke}--\eqref{ake} in terms of the photon density. 
Assuming that photons are confined to a box of volume $V$ with periodic boundary conditions, the density of states is
\[
g(\vb{k}) \dd^3 \vb{k} = \frac{2V}{(2\pi)^3} \dd^3 \vb{k} = \frac{2V}{(2\pi)^3} 
4\pi k^2 \dd k
\]
where $\vb{k}$ is the wave vector.  It is useful here to 
assume isotropy: in function of the dimensionless $x = \hbar \omega / k_B T$, the density of states is
\[
g(x) \dd x = \frac{2V}{(2\pi)^3} \left( \frac{k_B T}{\hbar c} \right)^3 4\pi 
x^2 \dd x
\] 
From here we can integrate the photon occupation number to get the total number of photons
\[ N_t = \int_0^\infty \dd{x} g(x) n(t,x)
 \]
That allows to write the spectral probability density
\begin{equation}\label{eq:spd}
\rho(t,x) = \frac{V}{N_t} \frac{1}{\pi^2} \left(\frac{k_B T}{\hbar c}\right)^3 
x^2 n(t,x) = \frac{x^2 n(t,x)}{2\zeta(3) Z_t},
\end{equation}
where $2\zeta(3) = \int_0^\infty \dd x \, x^2/(e^x-1) \simeq 2.404$.
By using the normalization of the spectral probability density while integrating \eqref{eq:spd} over $x$, we can write
\begin{equation}\label{zy}
Z_t =  \frac 1{2\zeta(3)}\,\int_0^\infty \dd x\, x^2 n(t,x) \propto  \frac{N_t}{V} \left(\frac{\hbar c}{k_B T}\right)^3
\end{equation}
which is a possibly time-dependent parameter, depending on the temperature and 
proportional to the number of photons per volume. Since the Kompaneets equation (still without reaction mechanism)
is photon-number preserving, $Z_t=Z$ is time-independent and yields for the 
stationary $n_\mu$ in \eqref{cc},
\[
Z = \frac{\text{Li}_3 (e^{\mu})}{\zeta(3)}
\]
with Li$_3$ the polylogarithm of order 3.
In that notation, the Kompaneets equation \eqref{ake} becomes
\begin{equation}\label{kp}
\frac{\partial \rho}{\partial t} (t,x) = -\frac{\partial}{\partial x}\left[\left(4x- x^2\left(1+2\zeta(3) Z\,\frac{\rho(t,x)}{x^2}\right)\right)\rho(t,x)\right] + \frac{\partial^2}{\partial x^2}\left[x^2 \rho(t,x)\right]
\end{equation}
Observe still that  $Z$
can be interpreted as the ratio of the actual photon spectral density to that 
of the Planck distribution corresponding to the same temperature.  For the 
Bose-Einstein distribution $n(t,x)= n_\text{BE}(x) = n_0(x) = 1 / (e^x - 1)$, 
the photon number equals
\[
N_\text{BE} = 2 \zeta(3) \frac{V}{\pi^2} \left( \frac{k_B T}{\hbar c} \right)^3
\]
corresponding to a spectral probability density with $Z = 1$:
\begin{equation}\label{eq:bespd}
\rho_\text{BE}(x) = \frac{x^2 n_{BE}(x)}{2\zeta(3)}
\end{equation}
We make a ``hot'' Planck spectral density by changing $x\rightarrow x/2$ (doubling the temperature), taking $n(x) = n_\text{BE}(x/2)$, leading to
\begin{eqnarray}\label{hott}
\rho_\text{BE}^\text{hot}(x) = \frac{x^2 n_\text{BE}(x/2)}{16\zeta(3)}
\end{eqnarray}
in which case $Z=8$.  On the other hand, as $Z$ goes to zero, we recover the Wien expression $\rho_\text{Wien}(x) = x^2 e^{-x}/2$ to be the stationary spectral probability density for equation \eqref{kp} which becomes linear at $Z=0$.\\

As a final comment we note that the Kompaneets equation is positivity preserving \cite{positivity}. That is important because any solution of the equation will retain its sign, suggesting indeed to interpret this equation as of Fokker-Planck type. 

\section{Mesoscopics of Compton scattering: Kompaneets stochastic 
process}\label{ksp}	

When considering a single Compton scattering event, the transition rates are fully determined by the incident electron and photon momenta together with energy-momentum conservation. However, in the presence of repeated scattering, treating the electron as a classical particle and the photon as a boson, the transition rates must account for stimulated emission, \textit{i.e.}, transitions are enhanced if photons are present in the final state. Mathematically, we work on a symmetrized Fock space  \cite{kadanoff}, where transition rates (between incoming  $|\text{i}\rangle$ and final states $|\text{f}\rangle$) have an additional term coming from the matrix element
\begin{equation}
\left|\langle \text{f}\,| a^\dagger_{{\mathbf k'}}a_{\mathbf k}|\text{i}\rangle\right|^2 = (1+n({\mathbf k'}))\,n({\mathbf k})
\end{equation}
where $a^{(\dagger)}_{\mathbf k}$ are the annihilation (creation) operators in the Fock space  related to the photon momentum $\vb{k}$, and $n({\mathbf k})$ are occupation numbers.  That is the only ``interaction'' between the photons that we take into account so far.

\subsection{Jump process in reciprocal space}

As introduced in \cite{paper}, we start from a random walk of bosons in reciprocal space. Here we consider a bath of a large number $N$ of photons, confined to a cube of sidelength $\ell$ with 
periodic boundary conditions. We will end up working in the thermodynamic limit for fixed density $N/\ell^3$.  For now however the modes are quantized, at a fixed distance $\delta = 2\pi / \ell$.  We designate each photon by its wave vector $\vb{k}_i$ so that the full state of the bath is described by $\vb{K} = (\vb{k}_1, \dots, \vb{k}_N)$.  Yet, we must treat the photons indistinguishably, and soon we will work with occupation numbers.\\
Transitions $\vb{K}\rightarrow \vb{K}'$ between  states are restricted to 
these for which only one photon jumps to another wave vector, {\it i.e.}, to transitions
\[
\vb{K} = (\vb{k}_1, 
\dots,\vb{k}_i,\ldots, \vb{k}_N)\longrightarrow \vb{K}' = (\vb{k}_1, 
\dots, \vb{k}_i+\vb{a} \delta,\dots, \vb{k}_N)
\] 
where $\vb{a}$ is one of the six unit vectors. The corresponding rate of such a transition is of the form
\[
w_i(\vb{K}, \vb{a}\delta) = w(\vb{k}_i,\vb{a}\delta) \left(1 + n_{\vb{k}_i+\vb{a} \delta}(\vb{K} )\right)
\]
where
\[
n_{\vb{k}}(\vb{K}) = \sum_{i} \delta_{ \vb{k}, \vb{k}_i}
\]
(with the Kronecker delta) counts the number of photons at $\vb{k}$.  It realizes the stimulated emission in the process.  The rest of the rates is taken as usual,
\[
w(\vb{k},\vb{a}\delta) = D\left( \vb{k} + \frac{\vb{a} \delta}{2} \right)
\exp\left\{-\frac{\beta}{2}\left(U(\vb{k} + \vb{a} \delta) - 
U(\vb{k})\right)\right\}
\]
to satisfy detailed balance with energy function $U$ at inverse temperature $\beta$.   We also added a (time-symmetric) reactivity $D$, also to be specified below in the case of Compton scattering.\\

The thus defined process $\vb{K}(t)$ is Markovian and has backward generator $L_\delta$, to be applied to observables
$F(\vb{K})$, given by
\begin{equation}\label{ld}
L_\delta F(\vb{K})= \sum_{i,\vb{a}} w_i(\vb{K},\vb{a}\delta) \left(F(\vb{K}') 
-F(\vb{K})\right)
\end{equation}
where $\vb{k}'_j = \vb{k}_j$ for $j \neq i$ and 
$\vb{k}'_i = \vb{k}_i + \vb{a} \delta$.
So far, the process can be seen as a generalized zero range process \cite{blythe}.  That generalization is sometimes referred to as a ``misanthrope process'', motivated by the convenience of monotonicity; see \cite{cocozza,sethuraman}.  Here that name is less appropriate as the dependence on the target configuration is one of ``stimulation''. \\
We are not staying with the multiparticle dynamics generated by \eqref{ld}, as we wish to find the dynamics for a single tagged photon.  To have a rigorous understanding of the dynamics of a tagged particle in a ``misanthrope'' (or stimulated) zero range process is far from trivial; see also \cite{jara}.  Our approach will therefore be more heuristic.\\

To start, we find the time evolution of the expected occupation numbers by applying the above rule to the observables
$n_{\vb{k}}$ and by noting that  \[ n_{\vb{k}}(\vb{K}') - n_{\vb{k}}(\vb{K}) = \delta_{\vb{k},\vb{k}_i + \vb{a} 
	\delta} - \delta_{\vb{k},\vb{k}_i} \]
Hence,
\begin{alignat}{3}
L_\delta n_{\vb{k}} (\vb{K}) &=& \sum_{i,\vb{a}} w(\vb{k}_i,\vb{a}\delta) \left(1 + 
n_{\vb{k}_i+ \vb{a}\delta}(\vb{K}) \right)& (\delta_{\vb{k},\vb{k}_i + \vb{a} 
	\delta} - \delta_{\vb{k},\vb{k}_i} )\nonumber\\
&=& \sum_{\vb{a}} w( \vb{k} - \vb{a}\delta, \vb{a}\delta)
\left(1 + n_{\vb{k}}(\vb{K}) \right) &\left(\sum_i \delta_{\vb{k},\vb{k}_i + 
	\vb{a} \delta}\right) \nonumber\\
&& & \  -  \sum_{\vb{a}} w(\vb{k}, \vb{a} \delta) \left(1 + n_{\vb{k}+ \vb{a}\delta}(\vb{K}) 
\right) \left(\sum_i \delta_{\vb{k},\vb{k}_i} \right)\nonumber\\
&=& \sum_{\vb{a}}  w( \vb{k} - \vb{a}\delta, \vb{a}\delta)
\left(1 + n_{\vb{k}}(\vb{K}) \right)& n_{\vb{k} - \vb{a} \delta}(\vb{K}) - w(\vb{k}, \vb{a} \delta) \left(1 + n_{\vb{k}+ \vb{a}\delta}(\vb{K}) 
\right) n_{\vb{k}}(\vb{K})\label{deln}
\end{alignat}

Continuing with \eqref{deln} and writing $n(t,\vb{k}) = \left\langle n_{\vb{k}}(\vb{K}(t)) \right\rangle$ for the expectation value over the process at time $t$, we thus get
\begin{align*}
{\pdv{n}{t}} (t,\vb{k}) = \sum_{\vb{a}}&  w(\vb{k} - \vb{a} \delta, \vb{a}\delta) 
\left\langle (1 + n_{\vb{k}} ) n_{\vb{k} - \vb{a}\delta} \right\rangle 
-  w(\vb{k}, \vb{a}\delta) \left\langle (1 + n_{\vb{k}+\vb{a} \delta} ) 
n_{\vb{k}} \right\rangle
\end{align*}
Next we assume that the correlations between occupations factorize.  That is not only part of standard kinetic theory; it originates mainly in the extremely weak interaction between photons.  We end up then with
\begin{align}
{\pdv{n}{t} }(t,\vb{k}) 
= \sum_{\vb{a}}  w(\vb{k} - \vb{a} \delta, \vb{a}\delta) (1 
+ n(t,&\vb{k})) n(t,\vb{k} - \vb{a}\delta) \nonumber \\
&-  w(\vb{k}, \vb{a}\delta) (1 + n(t,\vb{k}+\vb{a} \delta)) n(t,\vb{k})
\end{align}
which has the form of a nonlinear Master equation.  We already note the resemblance with the Kompaneets equation \eqref{ke} but that can be made more complete by taking the diffusion limit.

\subsection{Diffusion limit}

The process above can be considered in the limit $\delta\downarrow 0$ while rescaling time.
To obtain the limiting diffusion process we calculate the limiting backward generator $L = L_\delta/\delta^2$ on permutation-invariant functions $F$ which are piecewise constant on cubes of side $2\delta$ around $\vb{K}\in (\delta\bbZ)^{3N}$.   We remember from \eqref{ld} that
\begin{eqnarray}
L_\delta F(\vb{K}) = \sum_{i,a} D\left( \vb{k}_i + \frac{\vb{a} \delta}{2} \right) 
\exp\left\{-\frac{\beta}{2}\left(U(\vb{k}_i + \vb{a} \delta) - 
U(\vb{k}_i)\right)\right\}&&(1+ n_{\vb{k}_i+ \vb{a}\delta}(\vb{K}))\nonumber\\
&&\left(F(\vb{K}') 
-F(\vb{K})\right)
\end{eqnarray}
For expanding that to order $\delta^2$ we must take into account
\begin{align*}
&w(\vb{k},\vb{a}\delta) = \big(D(\vb{k}) + \frac{ \delta}{2}\vb{a}\cdot 
\nabla_{\vb{k}} D (\vb{k})\big)\;\big(1-\frac{\beta\delta}{2}\vb{a}\cdot 
\nabla_{\vb{k}} U (\vb{k})\big) \\
&n_{\vb{k+a\delta}} = n_{\vb{k}} +  \delta\,\vb{a}\cdot \nabla_{\vb{k}} n
\end{align*}
Whence, to nonvanishing order,
\begin{align}
\frac 1{\delta^2}L_\delta F(\vb{K}) = \sum_{i}D(\vb{k}_i)(1+ 
n_{\vb{k}_i}(\vb{K}))\Delta_{\vb{k}_i}&F(\vb{K})+\sum_{i,\vb{a}}\bigg{\{}\left(\frac{1}{2}\vb{a}\cdot \nabla_{\vb{k}} D (\vb{k}_i) 
-\frac{\beta}{2}D(\vb{k}_i)\vb{a}\cdot \nabla_{\vb{k}} U 
(\vb{k}_i)\right)\nonumber\\
&(1+n_{\vb{k}_i}(\vb{K}))\
+ D(\vb{k}_i) \,\vb{a}\cdot 
\nabla_{\vb{k}} n\bigg{\}}\vb{a}\cdot\nabla_{\vb{k}_i}F(\vb{K})\label{2o}
\end{align}
We should remember that $F$ depends on $\vb{K}$ only through the occupations $n_{\vb{k}}(\vb{K})$.
Moreover we are interested in the dynamics of a tagged photon, which amounts to 
a single-particle description.  We take therefore functions $F(\vb{K}) = 
\int\dd^3\vb{k} f(\vb{k}) n_{\vb{k}}(\vb{K})$ for the field $n_{\vb{k}}(\vb{K}) 
= \sum_j \delta(\vb{k}_j-\vb{k}) $,
\begin{equation}\label{sumj}
F(\vb{K}) =
\sum_j \int\dd^3\vb{k} f(\vb{k}) \, \delta(\vb{k_j}-\vb{k}) = \sum_j f({\vb{k}_j})
\end{equation}
In that case, expression \eqref{2o} can be rewritten line per line and per wave 
vector by using $\frac 1{\delta^2}L_\delta F(\vb{K}) = \int \dd^3\vb{k} 
\,n(\vb{k})\,{\cal L}f(\vb{k})$ for 
\begin{align}
{\cal L}f(\vb{k})= D(\vb{k})(1+ n_{\vb{k}})&\Delta_{\vb{k}}f(\vb{k}) +\sum_{\vb{a}} \bigg{\{}\frac{1}{2}\vb{a}\cdot \nabla_{\vb{k}} D (\vb{k})(1+n_{\vb{k}})\nonumber\\
&-\frac{\beta}{2}D(\vb{k})\vb{a}\cdot \nabla_{\vb{k}} U (\vb{k})(1+n_{\vb{k}})+ D(\vb{k}) \,\;\vb{a}\cdot \nabla_{\vb{k}} n\bigg{\}}\quad\vb{a}\cdot\nabla_{\vb{k}}f(\vb{k})\label{4o}
\end{align}
 Note now that
\begin{align}
\int\dd^3\vb{k}\,n_{\vb{k}}\,\bigg\{ D(\vb{k})\,n_{\vb{k}}\Delta_{\vb{k}}&f(\vb{k})\nonumber\\
&+ \sum_a\left[\frac{1}{2}\vb{a}\cdot \nabla_{\vb{k}} D (\vb{k})\,n_{\vb{k}}
+ D(\vb{k}) \,\;\vb{a}\cdot \nabla_{\vb{k}} n\right]\quad\vb{a}\cdot\nabla_{\vb{k}}f(\vb{k})\bigg\}=0
\end{align}
by partial integration.  We thus get
\begin{align}
\int\dd^3\vb{k}\,n(\vb{k})\,{\cal L}f(\vb{k})= &D(\vb{k})\,\Delta_{\vb{k}}f(\vb{k})\nonumber\\
&+\sum_a \left\{\frac{1}{2}\vb{a}\cdot \nabla_{\vb{k}} D (\vb{k})
-\frac{\beta}{2}D(\vb{k})\vb{a}\cdot \nabla_{\vb{k}} U (\vb{k})(1+n_{\vb{k}})\right\}\;\vb{a}\cdot\nabla_{\vb{k}}f(\vb{k})\label{5o}
\end{align}
Remember that we interpret $n_{\vb{k}}$ as a given continuum field.  As a consequence, the limiting diffusion of the tagged photon is given by
 \begin{equation} \label{3dsde}
\dot{\mbf{k}}_t = \nabla_{\vb{k}}{D}({\mbf{k}_t}) -\beta D(\vb{k}_t)\,\nabla_{\vb{k}} U (\vb{k}_t)(1+n_{\vb{k}_t}) + \sqrt{2D(\mbf{k}_t)}\,\mbf{\Xi}_t	
\end{equation}
where $\mbf{\Xi}_t$ is a standard white noise on $\bbR^3$ and It\^o-convention must be applied.  The derivation of \eqref{3dsde} from \eqref{5o} is not trivial.  It would follow standard practice without the nonlinearity (stimulated emission) by understanding the backward generator $\cal L$ as the generator of the time-evolution of the walk in reciprocal space.  Yet, full rigor is not attempted here and the precise derivation of the tagged photon dynamics remains mathematically challenging; see also \cite{fra} for more details. Numerical checks will follow below.

	\subsection{Tagged photon nonlinear Markov process}
We get to our main result.
Assume that the process is isotropic to suppose that $D$ and $U$ only 
depend on the radial component $x$ with ${\mbf{k}} = x/(\beta\hbar 
c)\mbf{\hat{x}}$.  We keep referring to $x = \beta \hbar \omega$ as the dimensionless frequency.\\
We apply It\^o's lemma to \eqref{3dsde}, with calculation in Appendix \ref{itoder}, to end up with
\begin{equation} \label{kp-ito2}
\dot x	=  2\frac{{D}(x)}{x}+ \partial_xD(x) - D(x) \partial_x U(x)(1+n(t,x))  + \sqrt{2{D}(x)}\, \xi_t
\end{equation}
where $\xi_t$ is again standard white noise.  That It\^o-stochastic process \eqref{kp-ito2} is what we call the Kompaneets process, and its construction and simulation (in the next Section) is our first main result.  It is the (nonlinear) Langevin dynamics associated to the Kompaneets equation \eqref{kp} in the case where 
\begin{equation}\label{uu}
U(x) = x,\qquad D(x) = x^2
\end{equation}
Note the dependence on the photon occupation field $n(t,x)$.  What we have here is a mean-field nonlinear Markov process for the tagged photon, 
where the field $n(t,x)$  represents the empirical occupations \cite{kolokoltsov, frank}; see also \cite{funaki}.  It 
realizes the Kompaneets equation  as a nonlinear Fokker-Planck equation.  That 
follows the spirit of the McKean-Vlasov equation \cite{mckean}, where the mean-field interaction leads to the nonlinearity in a multi-particle limit.  Our 
derivation has been ``theoretical'' but confirmation of the soundness of our 
approach is obtained in the next section.  One difficulty we ignore here is the 
singularity in  the modulus of the vector ${\mbf{k}}$.  It also implies that the 
white noise can lead to negative values of $x$.  Yet, in the simulations when starting with a particle number less than or equal to the value 
corresponding to the Planck distribution ($Z\leq 1$), even without an explicit 
boundary condition, there is no probability flux through the origin, that is, no negative frequencies are observed.  For a sufficiently small time step in the simulation, no particle ever reaches $x=0$.

\section{Simulation of the Kompaneets process}\label{simul}
Suppose the following It\^o stochastic differential equation for $X_t\in \bbR$,
\begin{equation}\label{genp}
	\dot {X}_t = B(X_t)  +  \sqrt{2D(X_t)}\;\xi_t,\qquad t\geq 0
\end{equation}

with standard white noise $\xi_t$. Then, the Euler-Maruyama algorithm reads
\begin{equation*}\label{em}
X_{t_{i+1}}=X_{t_i} + B(X_{t_i})\Delta t + \sqrt{2D(X_{t_i})} \Delta W_i + 
O(\Delta t^{3/2})
\end{equation*}
in which $\Delta t = t_{i+1} - t_i$ is the timestep and $\Delta W_i$ is a 
Gaussian random variable with mean $0$ and variance $\Delta t$ at iteration 
$i$.\\
If $B = B(X_t,\rho_t)$ depends also on the distribution function $\rho_t$  at 
time $t$, we need to consider an ensemble of processes. Here we take $N_E$ 
independent Kompaneets processes, and approximate the probability density 
$\rho_t$ with the empirical density $\rho_t^E$, which follows from a histogram 
of the frequencies of the ensemble. We expect in the limit $\rho_t^E \to 
\rho_t$ (large $N_E$) that the implementation of stimulated emission is exact. 
Furthermore, a initial value for $Z$ is chosen, see \eqref{zy}, which can be interpreted as the 
ratio of the total number of photons to the number of photons which would be 
present in the same volume if the occupation number were to follow a 
Bose-Einstein distribution. We can then at any point in time invert 
\eqref{eq:spd} yielding the empirical occupation number
\begin{equation*}
n_t^E(x) = 2\zeta(3) Z \frac{\rho_t^E(x)}{x^2}.
\end{equation*}
\\
 We simulate an ensemble of around $N_E\simeq 10^5$ particles, fixing the total 
time $\simeq 100$ and using a timestep of $\dd t = 10^{-4}$. The timestep was 
chosen to be low enough that lowering it did not appreciably change the 
simulation. The histograms are binned with a step of $\dd x = 0.05$ in 
frequency domain. As most of the distribution is concentrated between 0 and 10, 
this implies a bin will have on average about \num{500} particles.

The Euler-Maruyama algorithm for each particle reads at each iteration $i$
\begin{equation}\label{emalg}
x_{t_{i+1}}=x_{t_i} + \left(4x_{t_i} - x^2_{t_i}\left(1+n^E_{t_i}(x_{t_i})\right)\right)\dd t + x_{t_i}\sqrt{2\dd t} 
u_i
\end{equation}
The above equation corresponds to the choices 
\begin{align}
	&B(x,\rho_t) =4x - x^2(1+n_t(x))&\text{with}&&n_t(x) = 2\zeta(3)Z\frac{\rho_t(x)}{x^2}
	\label{drift}\\ 
	&D(x) =x^2 &&&\label{diffusion}
\end{align}
for drift and diffusion in \eqref{genp}, respectively (equivalent of \eqref{uu}).  Note here that at low frequencies, $x\ll 1$, there is almost no activity when also $x^2 n_t(x)\ll 1$. 
At each iteration $i$, we update the frequency according to \eqref{emalg}. Note 
that we check at each step the histogram at the frequency obtained in the 
previous time step. Then, after the Euler-Maruyama step, the histogram is 
updated accordingly.
It is also important to mention here that we need to specify the initial occupation field $n_0(x)$, which determines the probability density $\rho_0(x)$ from which the initial $x_0$ is drawn.  The step from $n_0$ to $\rho_0$ requires specifying $Z$ in \eqref{eq:spd}.  In other words, there can be two different evolutions even when the initial $\rho_0$ are identical, if there is a different $Z$.\\

\begin{figure}
	\includegraphics[width=0.49\textwidth]{{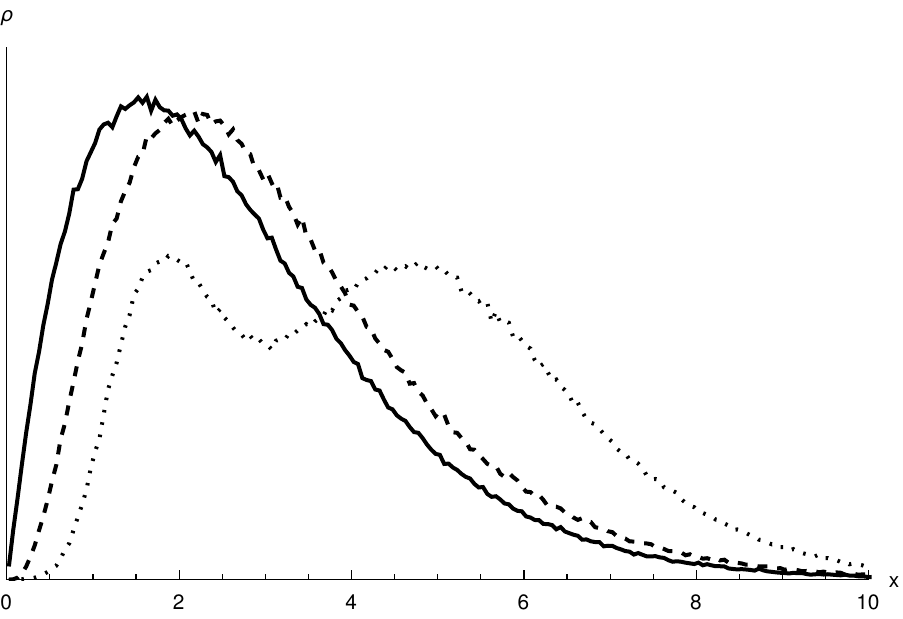}}\hfill%
	\includegraphics[width=0.49\textwidth]{{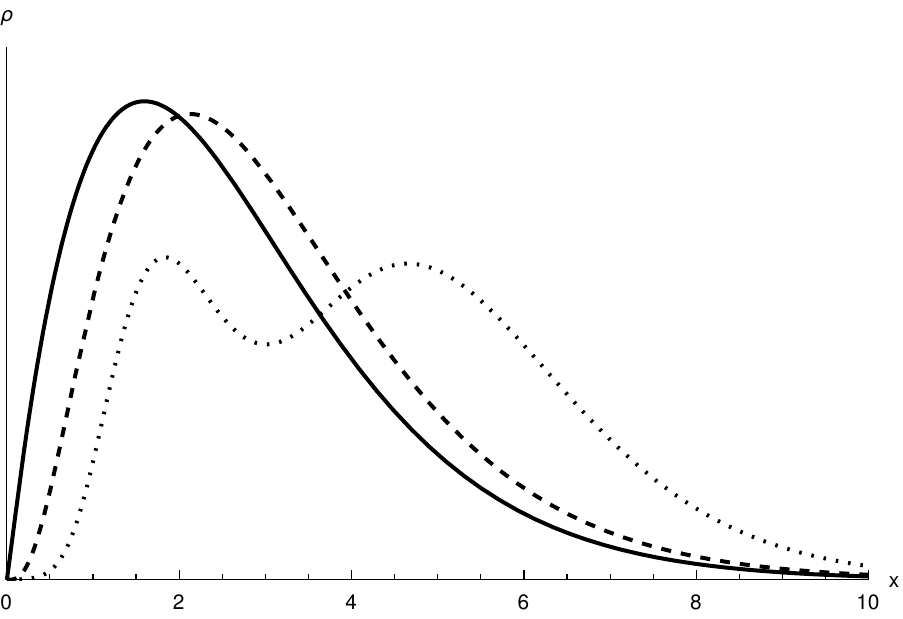}}
	\caption{The spectral probability density $\rho_t$ in function of the 
	dimensionless frequency $x$ with plots for $t=0.05,0.5$ and $\num{5e0}$, 
	respectively dotted, dashed and solid. We compare the Kompaneets process 
	\eqref{kp-ito2} on the left with a numerical solution of the Kompaneets 
	equation on the right. The initial condition (not shown here) was taken to 
	be a sum of two Gaussians centered around $x=2$ and $x=6$ with standard 
	deviations of respectively 2 and 1, and weights of respectively $1/3$ and 
	$2/3$.}
	\label{fig:spd-compare}
\end{figure}

To validate the correctness of our (somewhat heuristic) derivation, we compare the empirical spectral probability density obtained by our process to a numerical integration of the Kompaneets process. The results are shown in fig.~\ref{fig:spd-compare} and show a clear agreement between the two. The initial condition as described in the caption, a sum of two Gaussians, is not supposed to be physical but was purposefully chosen to better visually highlight the accordance.


To quantitatively check how much the obtained histogram deviates from the 
theoretical prediction, we can for every time $t$ compute the excess 
$\rho_\text{NUM}(t,x) - \rho_\text{BE}(t,x)$, where $\rho_\text{NUM}(t,x)$ is 
the numerical integration of the Kompaneets equations. Our simulations show 
that over the entire frequency range the results have a precision on the order 
of \num{e-3}.

\begin{figure}
	\includegraphics[width=0.8\textwidth]{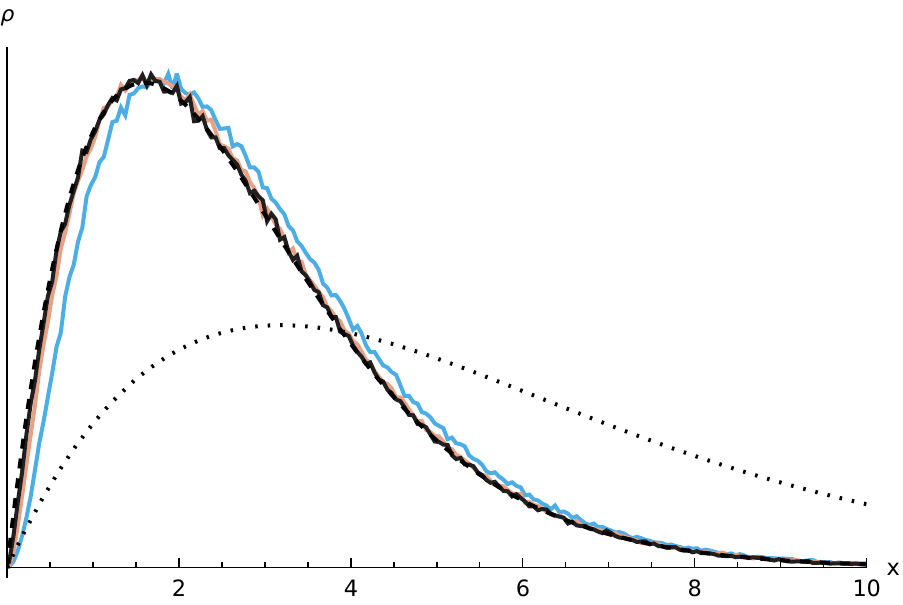}
	\caption{The spectral probability density $\rho_t$ in function of the dimensionless frequency $x$ for the Kompaneets process \eqref{kp-ito2}, with plots for $t=1,2,3$ going from right to left. The dotted line indicates the initial ``hot Planck'' condition corresponding to the occupation field $n_0(x) = n_\text{BE}(x/2)/8$ for which $Z=1$. Notice the fast thermalization to a Wien-like tail, followed by a slow final thermalization at lower frequencies ($x<6$).}
	\label{fig:spd-kompaneets}
\end{figure}

\begin{figure}
	\includegraphics[width=0.8\textwidth]{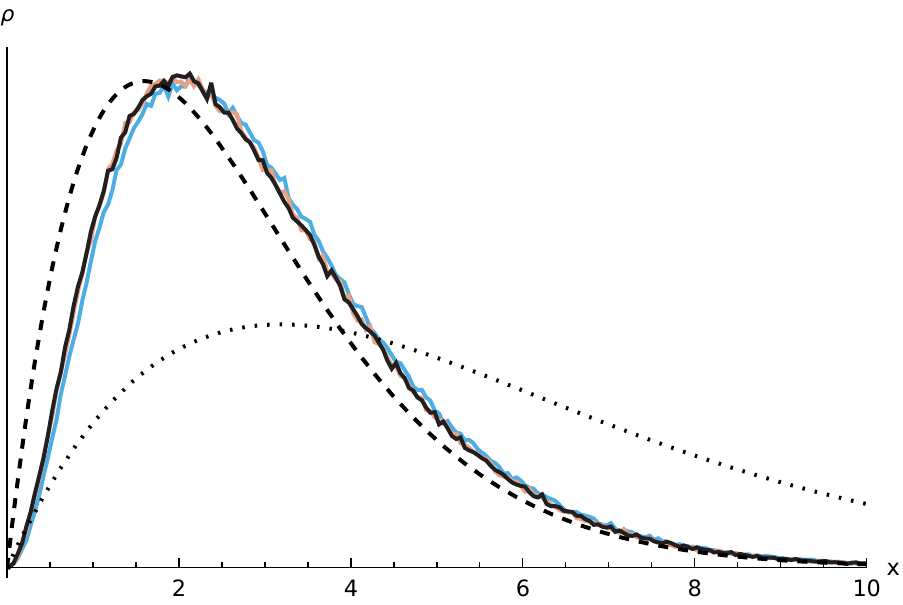}
	\caption{Similar to Fig.~\ref{fig:spd-kompaneets}, but with $Z=0.1$, {\it i.e.},  with the initial occupation field $n_0(x) = n_\text{BE}(x/2)/80$. There is a marked difference (going like $x^2$ at low frequency) with the dashed line representing the Planck density, as there is particle conservation. There is only the fast thermalization to a(n almost) Wien density because of the absence of low-frequency activity,  $x^2 n_t(x)\ll 1$ at low $x$.}
	\label{fig:spd-boltzmann}
\end{figure}

If we lower $Z$ to, say, $0.1$, the frequency spectral density instead 
converges to something closely resembling the Wien distribution $\simeq 
x^2e^{-x}$, as verified in Fig.~\ref{fig:spd-boltzmann}. This corresponds to 
\eqref{cc} with $\mu\simeq -2.13$. There is no further convergence to the 
Planck law here because there is particle conservation and because of the 
absence of low-frequency drift or diffusion for $x\ll 1$, making both 
\eqref{drift} and \eqref{diffusion} negligible since  $n_t(x)$ is of order 1 
there.\\

\begin{figure}
	\includegraphics[width=0.8\textwidth]{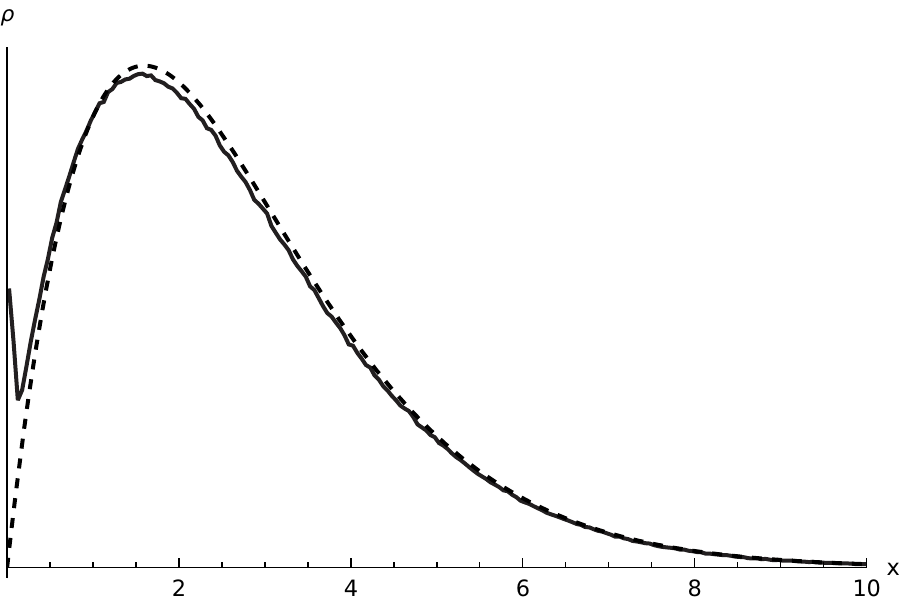}
	\caption{"Stationary" spectral density for $Z=1.1$. The dashed line is the 
	Planck density. Notice the formation of a condensate around the origin. In 
	that case, there is also the appearance of non-physical negative 
	frequencies. The condensate and the non-physical behavior may be remedied by 
	considering radiative processes like Bremsstrahlung.}
	\label{fig:spd-Zcondensate}
\end{figure}

We conclude from these simulations that the constructed process \eqref{kp-ito2}
can be reliably simulated and indeed, for its deterministic behavior reproduces 
the time-evolution \eqref{kp}, and hence the Kompaneets equation \eqref{ke}.  
We have given examples of various initial conditions and we conclude that the 
Kompaneets process shows two different time-scales: one is a fast relaxation to 
the Planck shape, better at high frequencies, after which a slower relaxation 
occurs shifting the distribution at the correct temperature.  Such a fast 
prethermalization followed by a slower adjustment is not uncommon in kinetic 
equations; it already happens in the classical Boltzmann equation where a local 
Maxwellian is rapidly established. \\
Finally, we note that when taking $Z>1$, a condensate is expected to form 
\cite{levermore}.  Fig.~\ref{fig:spd-Zcondensate} shows that condensate appearing for the simulated spectral density for $Z=1.1$. In that case, we also find particles 
going out of bounds at $x=0$, leading to unphysical negative energies. However, that may be
remedied by non-particle-conserving processes (e.g. double Compton or Brehmsstrahlung), which absorb the condensate at 
the origin (though simulation results are not shown here).

\section{Bremsstrahlung and double Compton scattering}\label{rbdc}
 
Double Compton scattering and  Bremsstrahlung are additional radiation processes, physically similar in controlling 
emission and absorption of photons. Since the number of photons is not conserved in this case, we can regard them as mechanisms to control the photon number density. \\
Both can be implemented as stochastic processes but we skip the derivations.  We refer to \cite{emrmodel, longair}, where the analysis that leads to the master equation was first obtained by Kompaneets (Eq.\;(18) in \cite{kompa}).\\

The Bremsstrahlung process resembles a chemical reaction, where photons are produced by electrons as they are deflected by nuclei. Here, we use the expression derived by Kompaneets to write in a slightly different manner
\begin{equation}
\left(\frac{\partial n}{\partial t}\right)_\text{B} = \frac{\tau_C}{\tau_B(x)} 
\left( \frac{1}{e^x-1} - n(t,x)\right)
\end{equation}
where we have used $\tau_C$, the characteristic Compton timescale of 
\eqref{shift}, to make the comparison with the frequency dependent timescale 
$\tau_B(x)$ of  Bremsstrahlung. Notice how there is no diffusion or drift, 
Bremsstrahlung instead provides a pointwise convergence of the occupation 
numbers towards the Bose-Einstein distribution.\\
If we temporarily reinstate units, and assuming for a moment that all nuclei are single protons, which means their 
density matches that of the electrons, we find that
\begin{equation}\label{brem}
\frac{\tau_C}{\tau_B(x)} = \frac{9\alpha n_e}{8} \sqrt{\frac{\pi}{2} 
\left(\frac{m_e c^2}{k_B T}\right)^3} \left( \frac{\hbar c}{k_B T}\right)^3 
\frac{\sinh(x/2) K_0(x/2)}{x^3}
\end{equation}
where $\alpha$ is the fine structure constant and $K_0$ is a modified Bessel 
function of the second kind \cite{macdon} (adapted to our notation from \cite{kompa}). Given 
the $x^{-3}$ dependence of the timescale, we see that Bremsstrahlung dominates 
the very lowest frequencies, but is negligible everywhere else.  Qualitatively, they reproduce the above reactive processes.\\

Similarly, we can treat the double Compton mechanism, whose radiative nature makes the treatment parallel to Bremsstrahlung. We choose to not give the detailed expression for the timescale, but we follow \cite{lightman} to note that double Compton scattering is faster in producing the Planck spectrum. In fact, both have nearly the same frequency dependence, but Bremsstrahlung is more relevant for rarefied photon gases and matter-dominated plasmas. Conversely, double Compton scattering becomes more important at higher electron-temperatures and for radiation-dominated plasmas \cite{lightman}. \\

 Nevertheless, in our simulation of the extened Kompaneets process below, we do not need the detailed implementation of the 
processes.  Using the fact that both radiative mechanisms have 
``nearly'' the same frequency dependence, we control in the 
simulations the number of photons by hand.   Both processes completely 
dominate the lower frequency ranges, but are negligible compared to Compton 
scattering for higher frequencies. In the simulations of the Kompaneets process we will simply add or remove particles under a 
certain cutoff such that their occupation numbers always exactly correspond to 
the Bose-Einstein distribution for a given volume and temperature.

\section{Extension of the Kompaneets process}\label{extk-sec}

It is well known that the Boltzmann equation, the traditional point of departure to obtain \eqref{ke}, may wash out certain nonequilibrium degrees of freedom of the electron bath: any isotropic distribution \cite{barbosa,brown,brown2, peebles} for the bath (which respects a mild constraint on the decay rate towards zero at infinity)\cite{paper} yields the Kompaneets equation with a suitable redefinition of the temperature.\\
As proposed in \cite{paper}, the Kompaneets equation can be extended to include more general diffusivities and a possible driving.   In the present section we directly interfere with the Kompaneets process to add nonequilibrium features to the photon process, leaving aside for a moment the specific origin.  There will in fact be two main directions of nonequilibrium, effectively introduced in a generalized Kompaneets equation.\\

On the level of the Kompaneets equation, the nonequilibrium extension is taken to be
\begin{equation}\label{extk}
\frac{\partial n}{\partial t}=\frac{1}{x^2}\frac{\partial}{\partial x} x^4 \left\{  \frac{\partial n}{\partial x} (t,x) +  (1+bx^k)[1+n(t,x)]n(t,x)\right\} + \frac{c}{x^2}\frac{\partial}{\partial x} x \left\{  \frac{\partial n}{\partial x} (t,x)\right\}
\end{equation}
parametrized by positive constants $b,c$ and $k$.
As before, we have changed to dimensionless variables.  That makes a special case of a more general extended Kompaneets equation
\begin{equation}\label{aeke}
	x^2 \frac{\partial n}{\partial t}(t,x) = \frac{\partial}{\partial x} x^2 D(x)\left\{  \frac{\partial n}{\partial x} (t,x) +  g(x)[1+n(t,x)]n(t,x)\right\}
\end{equation}
for
 \begin{equation}\label{ext-dri}
D(x)= x^2 + \frac c{x},\qquad  g(x)=\frac{1 + bx^k}{1+cx^{-3}}
\end{equation}
Obviously, the case $c=b=0$ recovers the standard Kompaneets equation.  
The change in the drift where to the energy function $U(\omega) = \hbar\omega$ (for the standard Kompaneets process) an extra potential $\sim \omega^{k+1}$ is added, is motivated by the search for higher order drivings where the force  may depend on the frequency. One is to change the drift, where we imagine a driving that depends on the frequency instead of a constant ``force" as in the Kompaneets process.  Obviously, we do not assume photon-photon scattering, which is extremely weak in vacuum \cite{silveira}. The drift can however be realized by employing strong light-matter coupling, resulting in strong effective energy exchanges (and dissipation) on single-photon level \cite{roy2}.  In other words, we change \eqref{uu}  (or \eqref{drift}) and replace the constant $U'$ with a frequency-dependent force $g$.\\
A second  change is to add an extra diffusion (changing \eqref{diffusion}) where the diffusion constant no longer has the features of Compton scattering, but gets modified at low frequencies. By making $c\neq 0$ an extra diffusion has been added in frequency space where the diffusivity decays with frequency, certainly unlike Compton-type processes.  Obviously, the dependence $c/x$ in \eqref{ext-dri} should not be taken literally all the way to $x\downarrow 0$ but an appropriate cutoff will be installed in the simulation (in the next section). In general, while interested in low-frequency effects, we are not interested in the very-low frequency-behavior $x\downarrow 0$. The extra diffusive term (the last term in \eqref{extk}) was also introduced in \cite{arca}, motivated by hints of low-frequency modifications in the Planck law, as observed from ARCADE 2 data \cite{arcade1,arcade2} and EDGES experiment \cite{edges}. The idea is to drastically increase the activity of low-frequency photons, which are otherwise largely untouched in Compton scattering.  Or, in other words, to decrease the activity of high-energy photons.  The latter was motivated by a mechanism similar to stochastic turbulence.  One should think here of the analogue of creating suprathermal tails in electron velocity distributions by their Rutherford scattering against turbulent fields \cite{supra1,banerjee}.  From the Coulomb interaction, high energy electrons are least scattered.  Similarly, low-frequency photons are least affected by Compton scattering and that allows them to become abundant by nonequilibrium effects, breaking detailed balance with respect to the Planck law.\\

 A rearrangement is in order to rewrite \eqref{aeke} in terms of the spectral density $\rho(t,x)$, defined in the exact same way as \eqref{eq:spd}, yielding
\begin{equation}\label{sc}
	\frac{\partial}{\partial t} \rho(t,x) = - \frac{\partial}{\partial x} \left[ \left(2\frac{D(x)}{x} + \partial_xD(x) - D(x)g(x)\left(1+n(t,x)\right)\right)\rho(t,x)\right] + \frac{\partial^2}{\partial x^2}[D(x)\rho(t,x)]
\end{equation} 
Note the stimulated emission with the occupation number density $n(t,x)$; of course we can subsitute the spectral density from \eqref{eq:spd}.  Yet is numbers that matter, not probability.\\
Observe that \eqref{aeke} remains a continuity equation, photon-number preserving just like Kompaneets. It means that $Z_t$ in \eqref{eq:spd} for the density of the nonequilibrium extension \eqref{aeke} is also a time-independent prefactor, depending only on the environment temperature and the (constant) number of photons per volume.\\
Interpreting similarly as before the above equation as a Fokker-Planck type 
equation, we find the associated Langevin equation in the It\^o-interpretation
\begin{equation}\label{ekp}
	\dot{x} = 2\frac{D(x)}{x} + \partial_xD(x) - D(x)g(x)\left(1+n(t,x)\right) + \sqrt{2D(x)}\, \xi_t
\end{equation}
where $\xi_t$ is a standard white noise as before. According to \eqref{genp}, we have in this case
\begin{align}
B(x,\rho_t) = 2\frac{D(x)}{x} + \partial_xD(x) - D(x)g(x)\left(1+n(t,x)\right) 
\label{edrift}
\end{align}
with drift and diffusion given by expression \eqref{ext-dri}, and $n(t,x)$ derived from $\rho(t,x)$ as before.

To be compared with \eqref{kp-ito2}, the above equation defines the \textit{extended Kompaneets process} and makes our second main result: it is the fluctuating single-photon dynamics that takes into account physically interesting nonequilibrium features.  Note that a stationary solution to \eqref{extk} can be found, of the form
\begin{equation}\label{eq:stationary}
n_\text{st}(x;b,c;k) = \frac{1}{\exp\left(\int_{x_0}^x \dd s \, \frac{1+bs^k}{1+cs^{-3}}\right) - 1} 
\end{equation}
where $x_0$ is the cutoff for the extra diffusion, as we expect for a realistic situation. That solution is only valid for $x>x_0$. For $x\leq x_0$ the details of the cutoff should enter in the exponent as to make $n_\text{st}$ well defined. When adding radiative processes to \eqref{sc} at low frequencies such as Bremsstrahlung (see Section \ref{rbdc}), the stationary solution differs from \eqref{eq:stationary} as well.

\section{Simulation of the extended Kompaneets process}\label{simex}

Simulations of these extended processes are straightforward, we repeat the 
algorithm of Section \ref{simul}, but with different functions $B$ and $D$.  Furthermore, we perform an {\it ad hoc} implementation of Bremsstrahlung and double Compton scattering, by 
simply clamping the occupation numbers of the lowest frequencies to the Planck 
distribution, as explained in Section \ref{rbdc}. Specifically, we look at the 
lowest 1\% of the frequency range, which in our case consists of the lowest two 
bins, and compare the empirical occupation number with value expected from the 
Planck distribution. In case there is an excess, the surplus amount of particles 
are randomly chosen and removed from the simulation. On the other hand, if 
there is a deficit, particles are added, randomly distributed over the bin, to 
make up the difference.\\ 
Notice that as particles are being added to or removed from the ensemble, the 
value of $Z$ will change, i.e, it becomes time-dependent $Z=Z_t$.\\
In the following subsections we separate the influence of $b$ and $c$ in \eqref{ext-dri}; indeed we estimate it is interesting to study their influence separately and the physical mechanisms underlying their presence.

\subsection{Adding low-frequency diffusion}

We first take $c = 0.1$, leaving $b=0$ in \eqref{ext-dri}. A spurious 
difficulty arises for nonzero $c$, as the given diffusion and drift 
coefficients diverge near the origin. To remedy this, we implement a cutoff 
$x_0$ in the simulation, which we take to be $0.1$ (i.e.\;1\% of the relevant 
frequency range), where the extra diffusion term is taken to go linearly to 
zero as $x$ goes below this cutoff, as it is nonphysical for the activity to 
remain the same all the way to zero frequency anyway.

\begin{figure}
	\includegraphics[width=0.8\linewidth]{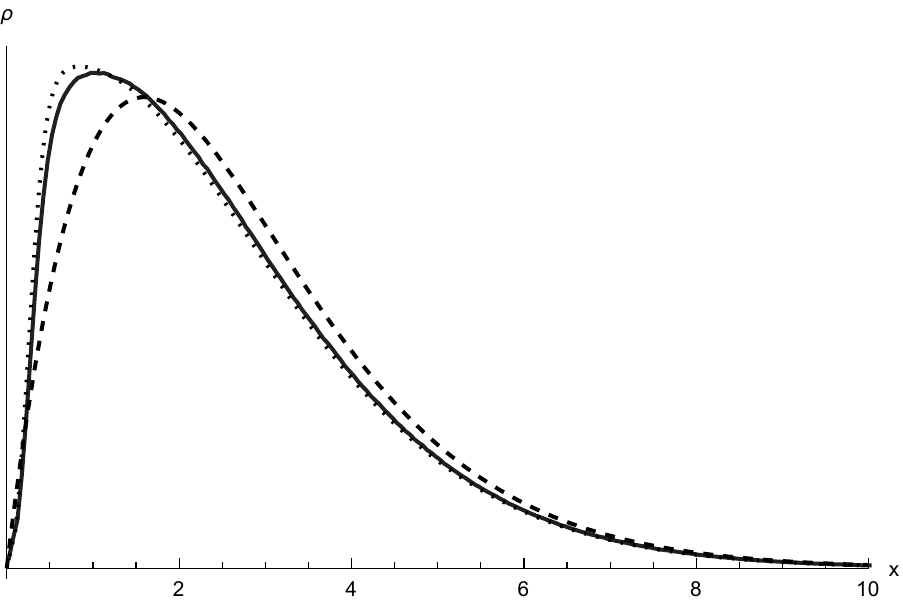}
	\caption{Solid line corresponds to the simulation of \eqref{ekp} using \eqref{ext-dri} for 
	$c=0.10, b=0$.  The dashed line is the Planck law. The dotted line depicts 
	the supposed stationary distribution \eqref{supsim}.
	}\label{fig:spd-c}
\end{figure}

Simulating with initially \num{2e5} particles, Fig~\ref{fig:spd-c} shows the 
resulting frequency histogram of the simulation for a total time $\simeq 5$ with 
the black solid line corresponding to (almost) stationarity. Deviations from 
the dashed Planckian distribution are evident, with an abundance of 
low-frequency photons, but keeping the Planck-shape at larger frequency.  
Together with the extra diffusivity, we implement reactive mechanisms. That 
makes the behavior below the cutoff frequency $x_0$ similar as if we were to 
turn off the extra diffusive term, i.e, to make $c=0$ for $x<x_0$. With that in 
mind we define\\
\[
h(x) = \int_0^x \dd x' \begin{cases}
\frac{1}{1+c / x'^3} & \text{for $x'>x_0$} \\
1 & \text{for $x' < x_0$}
\end{cases}.
\]
Then, the stationary distribution appears very well described as
\begin{equation}\label{supsim}
 n_\text{st}(x) = \frac{1}{\exp(h(x)) - 1} 
 \end{equation}
as is tested in Fig.~\ref{fig:spd-c}.  That theoretical density would 
correspond to $Z=1.82$, which is however not the value $Z\approx1.73$ that is 
found in the simulation. This is not surprising since the addition of reactive 
processes should make the stationary distribution slightly different at low 
frequencies.

\begin{figure}
	\includegraphics[width=0.8\linewidth]{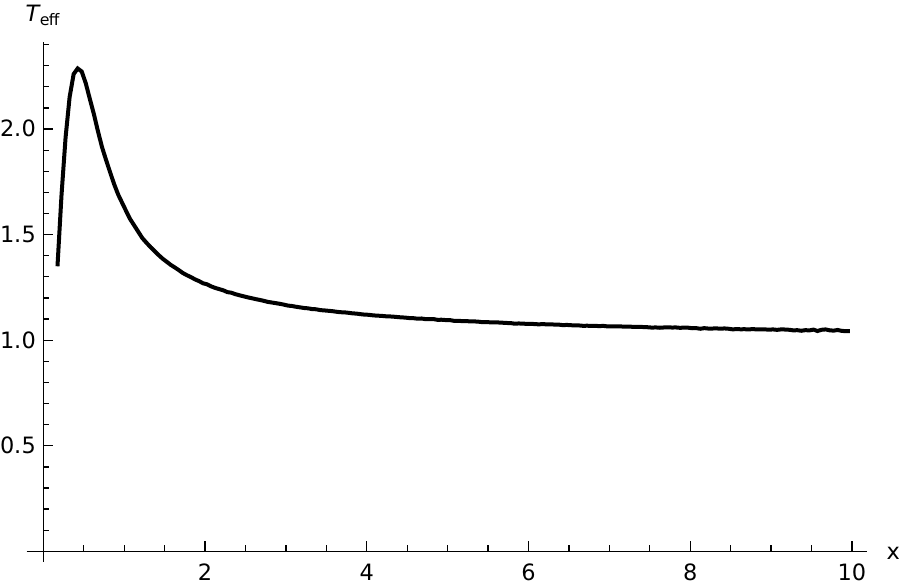}
	\caption{Profile of effective temperature corresponding to Fig. 4 and 
	defined in \eqref{effT}. The very-low frequency regime is dominated by the 
	{\it ad hoc} cutoff and procedures replacing Bremsstrahlung and double 
	Compton scattering, as explained in the text.  For low frequency $1<x<5$ 
	the effective temperature is significantly higher than the Planck-value.
	}\label{fig:T}
\end{figure}

We define an effective temperature $T_\text{eff}(x)$ as a function of the 
frequency, derived from the empirical occupation number at this frequency by 
inverting the Planck distribution,
\begin{equation}\label{effT}
n_{\text{Pl},T_\text{eff}}(x) = \frac{1}{\exp(x/T_\text{eff})-1}
\end{equation}

Because of our choice of units, this will be expressed as a ratio to the 
temperature of the background electron gas.  Fig.~\ref{fig:T} plots this 
effective temperature for the stationary distribution.

\subsection{Adding frequency-dependent drift}

Taking now a non-zero driving in \eqref{ext-dri}, we set $b=0.1$ and $k=0.4$, keeping $c=0$.  A marked deviation from the Planck law becomes visible.
It is also interesting to note a remarkable feature for $b\neq 0$, which is the formation of a condensate around $x=0$. However, the low-frequency reaction mechanism (either double Compton or Bremssthralung) is able to absorb the condensate. In fact, the appearance of that condensate can be explained by looking at the stochastic equation \eqref{ekp}, where the extra drift coming from $g(x)$ contributes in the stimulated emission term, making shifts towards negatives values of frequency more frequent.

\begin{figure}
	\centering
	\includegraphics[width=0.8\linewidth]{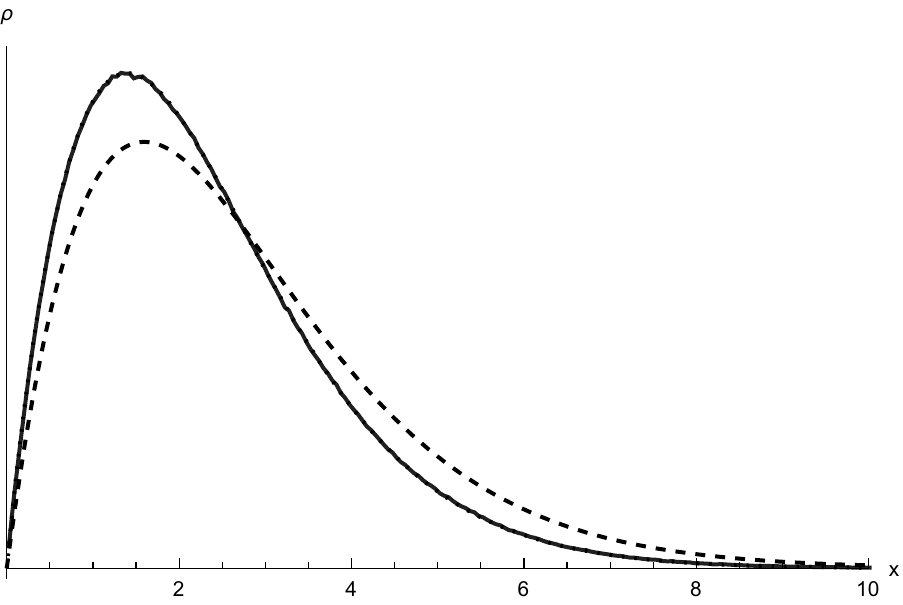}
	\caption{Stationary density profile when adding the frequency-dependent 
	drift with $b=0.1$ and $k=0.4$, keeping $c=0$ in \eqref{ext-dri}.  Again a 
	deviation is seen from the Planck law (dashed line), with the simulation 
	almost completely overlapping with the stationary solution.}
	\label{fig:spd-b}
\end{figure}
Here, in contrast with the diffusive addition in Fig.~\ref{fig:spd-c}, high frequencies get much more affected.

\section {Conclusions}\label{con}

The present paper achieves to construct photon processes in frequency space dominated by Compton scattering.  The stimulated emission produces the nonlinearities.  From the single-photon stochastic process, we enter the realm of time-dependent fluctuation phenomena for photon dynamics.  It implies we can start exploring nonequilibrium photon dynamics, much in the tradition of stochastic dynamics for open particle systems.   We have explored the addition of nonequilibrium drift and diffusion terms which affect mostly the low-frequency part in a modified Planck spectrum. \\

Various applications can be envisaged, subject of future studies, while the present paper sets the mathematical and simulational structure of the photon diffusion process. Despite the appearance of nonlinearities, due to stimulated emission, which makes a nontrivial aspect, we simulate this process using traditional algorithms for stochastic equations and conclude that relaxation to the equilibrium Planck law can be reliable simulated, as seen in Fig.~\ref{fig:spd-kompaneets}. The nonlinearity  is satisfactorily dealt by considering an ensemble of processes while using histograms to construct the empirical spectral density of photons at each time step. Physical processes such as Bremsstrahlung and double Compton can also be easily included by an \textit{ad hoc} procedure we describe in Section \ref{rbdc}. It means to continuously set, below some cutoff frequency, the density of photons to its corresponding value in the Planck spectrum, under the argument that low-frequency photons rapidly thermalize via reactive mechanisms.

The $Z$ parameter can be tuned as to make the stimulated emission weaker and 
convergence to a Wien-like spectrum is observed in the simulations; see 
Fig.~\ref{fig:spd-boltzmann}. In fact, this would correspond to a rarefied 
photon gas, where stimulated emission is expected to become less important. On 
the other hand, detailed balance with respect to Planck's law is broken by 
considering the extension to the Kompaneets equation. There, nonequilibrium 
drift and diffusion make extra contributions to the standard equation for the 
single-photon process. The proposed extension \eqref{extk} is parameterized by 
the values $b,c$ and $k$, such that $b=c=0$ corresponds to the standard 
Kompaneets process \eqref{kp-ito2}. The simulation results, 
Figs.~\ref{fig:spd-c} and \ref{fig:spd-b}, show that, for the values of 
parameters considered, modifications to the Planck law that ultimately lead to 
higher occupancy in the low-frequency part of the spectrum are seen. Those 
modifications match very well with the theoretical stationary solution 
\eqref{supsim}, but slight deviations are expected due to the inclusion of 
reaction mechanisms.\\
 From the mathematical point of view, the derivation of the Kompaneets process is interesting and challenging.  It appears to open a new theme in the domain of nonlinear Markov diffusions \cite{fra}.\\
Let us finish by emphasizing that the  proposed mechanisms remain conjectural and that the theory is largely phenomenological.  A follow-up can be found in \cite{res}, where a resetting mechanism is added to the Kompaneets process corresponding to random abrupt Doppler shifts to the low frequency regime. 

\begin{appendix}

\section{It\^o's lemma and the Kompaneets process}\label{itoder}

Suppose we have a (possibly) time-dependent function of the random variable $\vb{X}_t$
\[f=f(t,\vb{X}_t) \]
then, $f$ itself is a random variable. Suppose also that $\vb{X}_t$ follows the stochastic differential equation \begin{equation*}
	\dot{\vb{X}}_t=\vb{B}(\vb{X}_t) + \vb{Q}_t(\vb{X}_t)\vb{\Xi}_t
\end{equation*}
where $\vb{\Xi}_t$ is a three-dimensional white noise as before.

Then, It\^o's lemma states that $f$ satisfies the stochastic differential equation
\begin{equation}\label{ilem}
	\dot{f} (t, \vb{X}_t)= \frac{\partial f}{\partial t}+ \left(\nabla_{\vb{X}}f\right)\cdot\vb{B} +\frac{1}{2}\mathrm{Tr}\left[\vb{Q}^T_t\left(H_Xf\right)\vb{Q}_t\right] + \left(\nabla_{\vb{X}}f\right)^T\vb{Q}_t\, \vb{\Xi}_t
\end{equation}
where the positive $d\times d$ matrix $\vb{Q}_t$ is such that the diffusion tensor
\[\vb{D}=\frac{1}{2}\vb{Q}_t\vb{Q}^T_t\]
and $H_X$ is the Hessian matrix with respect to $\vb{X}_t$.

We use that to show in more detail the derivation of equation \eqref{kp-ito2}.  We begin with the $i$-th generator in \eqref{2o}
\begin{align}
	\frac 1{\delta^2}{L_i}_\delta F(\vb{K}) = D(\vb{k}_i)(1+ 
	n_{\vb{k}_i}(\vb{K}))\Delta_{\vb{k}_i}&F(\vb{K})+\sum_{\vb{a}}\bigg{\{}\left(\frac{1}{2}\vb{a}\cdot \nabla_{\vb{k}} D (\vb{k}_i) 
	-\frac{\beta}{2}D(\vb{k}_i)\vb{a}\cdot \nabla_{\vb{k}} U 
	(\vb{k}_i)\right)\nonumber\\
	&(1+n_{\vb{k}_i}(\vb{K}))\
	+ D(\vb{k}_i) \,\vb{a}\cdot 
	\nabla_{\vb{k}} n\bigg{\}}\vb{a}\cdot\nabla_{\vb{k}_i}F(\vb{K})\label{gen-i}
\end{align}
which leads by applying the steps \eqref{sumj}--\eqref{5o} to the It\^o process for the $i$-th photon
 \begin{equation} \label{i-3dsde}
 	\dot{\mbf{k_i}}_t = \nabla_{\vb{k}}{D}({\mbf{k_i}_t}) -\beta D(\vb{k_i}_t)\,\nabla_{\vb{k}} U (\vb{k_i}_t)(1+n_{\vb{k_i}_t}(\vb{K})) + \sqrt{2D(\mbf{{k_i}_t})}\,\mbf{\Xi}_t	
 \end{equation}

In order to simplify notation, we will opt to not differentiate the quantities from their dimensionless counterpart, therefore, it should be noted that all quantities appearing here are dimensionless. In particular, we shall identify the photon energy with its dimensionless version
\[\beta U (\omega_i) \to U(x_i)=x_i\]

We apply It\^o's lemma for 
\[f(\vb{k_i})=|\vb{k_i}|.\]

A simple calculation yields
\begin{align*}
	&\nabla_{\vb{k_i}}f =  \frac{\vb{k_i}}{|\vb{k_i}|}\\
	&H_{\vb{k_i}}f = 
	\begin{pmatrix}
		\frac{1}{|\vb{k_i}|}	-\frac{{k_{ix}}^2}{|\vb{k_i}|^3} & & \\
		&\frac{1}{|\vb{k_i}|} -\frac{{k_{iy}}^2}{|\vb{k_i}|^3}  &\\
		& & \frac{1}{|\vb{k_i}|}- \frac{{k_{iz}}^2}{|\vb{k_i}|^3}
	\end{pmatrix}
\end{align*}
recall from \eqref{i-3dsde} that
\begin{align*}
	&\vb{B}(\vb{k_i}) =\nabla_{\vb{k}}D(\vb{k_i}) - D(\vb{k_i})\nabla_{\vb{k}}{U}(\vb{k_i})(1+n_{\vb{k_i}}(\vb{K})) \\
	&\vb{Q}(\vb{k_i}) = \sqrt{2D(\vb{k_i})}\mathbf{I}_{3\times 3}
\end{align*}

Therefore, we can easily find all terms in It\^o's lemma \eqref{ilem}
\begin{align*}
	\begin{cases}
	&\partial_t f=0\\
	&\left(\nabla_{\vb{X}}f\right)\cdot\vb{B} =\frac{\vb{k_i}}{|\vb{k_i}|} \cdot\left(\nabla_{\vb{k}}D(\vb{k_i}) - D(\vb{k_i})\nabla_{\vb{k}}{U}(\vb{k_i})(1+n_{\vb{k_i}}(\vb{K}))\right)\\
	&\frac{1}{2}\mathrm{Tr}\left[\mbf{Q}^T_t\left(H_Xf\right)\mbf{Q}_t\right] = 2 \frac{D(\vb{k_i})}{|\vb{k_i}|}  \\
	&\left(\pmb{\nabla_X}f\right)^T\mbf{Q}_t  = \sqrt{2D(\vb{k_i})}\frac{\vb{k_i}}{|\vb{k_i}|}  
	\end{cases}
\end{align*}
we substitute back in \eqref{ilem} and write quantities in terms of the dimensionless frequency $\vb{k_i}=x_i\, \vb{\hat{x_i}}$. Repeating the process for $i=1,\dots, N$, taking the sum and the continuum limit for the generator in frequency space similarly than done in steps \eqref{2o}--\eqref{5o} we find \eqref{kp-ito2}.\\
Applying directly It\^o's lemma for a nonlinear Markov diffusion is not needed here, but is another mathematical question that is raised by the Kompaneets process.

\section{Data Availability Statement}

The data and software that support the findings of this study are openly available on GitLab\cite{gitlab}.

\end{appendix}

\bibliographystyle{ieeetr}
\bibliography{langevin-kompaneets}

\begin{thebibliography}{10}

\bibitem{koch}
J.~Koch, A.~A. Houck, K.~L. Hur, and S.~M. Girvin, ``Time-reversal-symmetry
  breaking in circuit-qed-based photon lattices,'' {\em Phys. Rev. A}, vol.~82,
  p.~043811, Oct. 2010.

\bibitem{fang}
K.~Fang, Z.~Yu, and S.~Fan, ``Realizing effective magnetic field for photons by
  controlling the phase of dynamic modulation,'' {\em Nature Photonics},
  vol.~6, pp.~782--787, Nov. 2012.

\bibitem{roushan}
P.~Roushan, C.~Neill, A.~Megrant, Y.~Chen, R.~Babbush, R.~Barends, B.~Campbell,
  Z.~Chen, B.~Chiaro, A.~Dunsworth, A.~Fowler, E.~Jeffrey, J.~Kelly, E.~Lucero,
  J.~Mutus, P.~J.~J. O'Malley, M.~Neeley, C.~Quintana, D.~Sank, A.~Vainsencher,
  J.~Wenner, T.~White, E.~Kapit, H.~Neven, and J.~Martinis, ``Chiral
  ground-state currents of interacting photons in a synthetic magnetic field,''
  {\em Nature Physics}, vol.~13, pp.~146--151, Oct. 2016.

\bibitem{pozar}
D.~Pozar, {\em Microwave Engineering}.
\newblock Wiley, 2004.

\bibitem{roy}
D.~Roy, ``Two-photon scattering by a driven three-level emitter in a
  one-dimensional waveguide and electromagnetically induced transparency,''
  {\em Phys. Rev. Lett.}, vol.~106, p.~053601, Feb. 2011.

\bibitem{metaphotonics}
A.~Baev, P.~N. Prasad, H.~Ågren, M.~Samoć, and M.~Wegener, ``Metaphotonics:
  An emerging field with opportunities and challenges,'' {\em Physics Reports},
  vol.~594, pp.~1--60, 2015.
\newblock Metaphotonics: An emerging field with opportunities and challenges.

\bibitem{supra1}
T.~Demaerel, W.~{De Roeck}, and C.~Maes, ``Producing suprathermal tails in the
  stationary velocity distribution,'' {\em Physica A: Statistical Mechanics and
  its Applications}, vol.~552, p.~122179, 2020.
\newblock Tributes of Non-equilibrium Statistical Physics.

\bibitem{banerjee}
T.~Banerjee, U.~Basu, and C.~Maes, ``Active velocity processes with
  suprathermal stationary distributions and long-time tails,'' {\em Phys. Rev.
  E}, vol.~101, p.~062130, June 2020.

\bibitem{fermi}
E.~Fermi, ``On the origin of the cosmic radiation,'' {\em Phys. Rev.}, vol.~75,
  pp.~1169--1174, Apr. 1949.

\bibitem{sturrock}
P.~A. Sturrock, ``Stochastic acceleration,'' {\em Phys. Rev.}, vol.~141,
  pp.~186--191, Jan. 1966.

\bibitem{brin}
C.~Univ, M.~Brin, W.~on~Dynamical~Systems, R.~Topics, K.~Burns, D.~Dolgopyat,
  and Y.~Pesin, {\em Fermi acceleration}, p.~149.
\newblock Contemporary mathematics - American Mathematical Society, American
  Mathematical Society, 2008.

\bibitem{dark}
T.~Buchert, ``{Dark Energy} from structure: a status report,'' {\em General
  Relativity and Gravitation}, vol.~40, pp.~467--527, Dec. 2007.

\bibitem{arcade1}
D.~J. {Fixsen}, A.~{Kogut}, S.~{Levin}, M.~{Limon}, P.~{Lubin}, P.~{Mirel},
  M.~{Seiffert}, J.~{Singal}, E.~{Wollack}, T.~{Villela}, and C.~A. {Wuensche},
  ``{ARCADE 2 Measurement of the Absolute Sky Brightness at 3-90 GHz},'' {\em
  \apj}, vol.~734, p.~5, June 2011.

\bibitem{arcade2}
M.~Seiffert, D.~J. Fixsen, A.~Kogut, S.~M. Levin, M.~Limon, P.~M. Lubin,
  P.~Mirel, J.~Singal, T.~Villela, E.~Wollack, and C.~A. Wuensche,
  ``{I}nterpretation of the {ARCADE} 2 absolute sky brightness measurement,''
  {\em The Astrophysical Journal}, vol.~734, p.~6, may 2011.

\bibitem{edges}
K.~Cheung, J.-L. Kuo, K.-W. Ng, and Y.-L.~S. Tsai, ``The impact of {EDGES}
  21-cm data on dark matter interactions,'' {\em Physics Letters B}, vol.~789,
  pp.~137--144, 2019.

\bibitem{arca}
M.~Baiesi, C.~Burigana, L.~Conti, G.~Falasco, C.~Maes, L.~Rondoni, and
  T.~Trombetti, ``Possible nonequilibrium imprint in the cosmic background at
  low frequencies,'' {\em Phys. Rev. Research}, vol.~2, p.~013210, Feb. 2020.

\bibitem{kompa}
A.~S. {Kompaneets}, ``{The Establishment of Thermal Equilibrium between Quanta
  and Electrons},'' {\em Soviet Journal of Experimental and Theoretical
  Physics}, vol.~4, pp.~730--737, May 1957.

\bibitem{levermore}
C.~D. Levermore, H.~Liu, and R.~L. Pego, ``Global dynamics of bose--einstein
  condensation for a model of the kompaneets equation,'' {\em SIAM Journal on
  Mathematical Analysis}, vol.~48, no.~4, pp.~2454--2494, 2016.

\bibitem{liedahl}
D.~A. {Liedahl}, {\em {The X-Ray Spectral Properties of Photoionized Plasma and
  Transient Plasmas}}, vol.~520, p.~189.
\newblock 1999.

\bibitem{blandford}
R.~D. Blandford and E.~T. Scharlemann, ``On induced {Compton} scattering by
  relativistic particles,'' {\em Astrophysics and Space Science}, vol.~36,
  pp.~303--317, Sept. 1975.

\bibitem{sunyaeveffect}
R.~A. Sunyaev and Y.~B. Zeldovich, ``Distortions of the background radiation
  spectrum,'' {\em Nature}, vol.~223, pp.~721--722, Aug. 1969.

\bibitem{sunyaev}
R.~A. {Sunyaev} and Y.~B. {Zeldovich}, ``{The Observations of Relic Radiation
  as a Test of the Nature of X-Ray Radiation from the Clusters of Galaxies},''
  {\em Comments on Astrophysics and Space Physics}, vol.~4, p.~173, Nov. 1972.

\bibitem{practical}
D.~G. Shirk, ``A practical review of the kompaneets equation and its
  application to {Compton} scattering,'' 2006.

\bibitem{gui}
G.~E. Freire~Oliveira, ``Statistical mechanics of the {Kompaneets} equation,''
  Master's thesis, KU Leuven. Faculteit Wetenschappen, Leuven, 2021.

\bibitem{zeldovich}
Y.~B. Zeldovich, ``Interaction of free electrons with electromagnetic
  radiation,'' {\em Soviet Physics Uspekhi}, vol.~18, pp.~79--98, Feb. 1975.

\bibitem{buet}
C.~Buet, B.~Despr{\'e}s, and T.~Leroy, ``{Anisotropic models and angular
  moments methods for the {Compton} scattering}.'' working paper or preprint,
  Feb. 2018.

\bibitem{pitrou}
C.~Pitrou, ``Radiative transport of relativistic species in cosmology,'' {\em
  Astroparticle Physics}, p.~102494, 2020.

\bibitem{barbosa}
D.~Barbosa, ``A note on {Compton} scattering,'' {\em The Astrophysical
  Journal}, vol.~254, pp.~301--308, 1982.

\bibitem{brown}
L.~S. Brown and D.~L. Preston, ``Leading relativistic corrections to the
  kompaneets equation,'' {\em Astroparticle Physics}, vol.~35, no.~11,
  pp.~742--748, 2012.

\bibitem{itoh}
N.~Itoh, Y.~Kohyama, and S.~Nozawa, ``Relativistic corrections to the
  {Sunyaev-Zeldovich} effect for clusters of galaxies,'' {\em The Astrophysical
  Journal}, vol.~502, no.~1, p.~7, 1998.

\bibitem{itoh2}
N.~Itoh and S.~Nozawa, ``Relativistic corrections to the {Sunyaev-Zeldovich}
  effect for extremely hot clusters of galaxies,'' {\em A\&A}, vol.~417, no.~3,
  pp.~827--832, 2004.

\bibitem{cooper}
G.~Cooper, ``Compton {Fokker-Planck} equation for hot plasmas,'' {\em Physical
  Review D}, vol.~3, no.~10, p.~2312, 1971.

\bibitem{kohyama1}
S.~Nozawa and Y.~Kohyama, ``Analytical study on the {Sunyaev-Zeldovich} effect
  for clusters of galaxies,'' {\em Physical Review D}, vol.~79, no.~8,
  p.~083005, 2009.

\bibitem{kohyama2}
S.~Nozawa, Y.~Kohyama, and N.~Itoh, ``Analytical study on the
  {Sunyaev-Zeldovich} effect for clusters of galaxies. ii. comparison of
  covariant formalisms,'' {\em Physical Review D}, vol.~82, no.~10, p.~103009,
  2010.

\bibitem{kohyama3}
S.~Nozawa and Y.~Kohyama, ``Relativistic corrections to the kompaneets
  equation,'' {\em Astroparticle Physics}, vol.~62, pp.~30--32, 2015.

\bibitem{paper}
G.~E. {Freire Oliveira}, C.~Maes, and K.~Meerts, ``On the derivation of the
  {Kompaneets} equation,'' {\em Astroparticle Physics}, vol.~133, p.~102644,
  2021.

\bibitem{positivity}
O.~Kavian, ``Remarks on the {Kompaneets} equation, a simplified model of the
  {Fokker-Planck} equation,'' in {\em Nonlinear Partial Differential Equations
  and their Applications - Coll{\`{e}}ge de France Seminar Volume {XIV}},
  pp.~469--487, Elsevier, 2002.

\bibitem{kadanoff}
L.~P. Kadanoff, {\em Quantum statistical mechanics}.
\newblock CRC Press, 2018.

\bibitem{blythe}
R.~A. Blythe and M.~R. Evans, ``Nonequilibrium steady states of matrix-product
  form: a solver{\textquotesingle}s guide,'' {\em Journal of Physics A:
  Mathematical and Theoretical}, vol.~40, pp.~R333--R441, Oct. 2007.

\bibitem{cocozza}
C.~Cocozza-Thivent, ``Processus des misanthropes,'' {\em Zeitschrift f{\"u}r
  Wahrscheinlichkeitstheorie und Verwandte Gebiete}, vol.~70, no.~4,
  pp.~509--523, 1985.

\bibitem{sethuraman}
S.~Sethuraman, ``On diffusivity of a tagged particle in asymmetric zero-range
  dynamics,'' 2004.

\bibitem{jara}
M.~Jara, C.~Landim, and S.~Sethuraman, ``Nonequilibrium fluctuations for a
  tagged particle in one-dimensional sublinear rate zero-range processes,''
  2010.

\bibitem{fra}
T.~D. Frank, {\em Nonlinear Fokker-Planck equations: fundamentals and
  applications}.
\newblock Springer Science \& Business Media, 2005.

\bibitem{kolokoltsov}
V.~N. Kolokoltsov, {\em Nonlinear {Markov} Processes and Kinetic Equations}.
\newblock Cambridge Tracts in Mathematics, Cambridge University Press, 2010.

\bibitem{frank}
T.~D. Frank, ``Strongly nonlinear stochastic processes in physics and the life
  sciences,'' {\em {ISRN} Mathematical Physics}, vol.~2013, pp.~1--28, Mar.
  2013.

\bibitem{funaki}
T.~Funaki, ``A certain class of diffusion processes associated with nonlinear
  parabolic equations,'' {\em Zeitschrift f{\"u}r Wahrscheinlichkeitstheorie
  und Verwandte Gebiete}, vol.~67, no.~3, pp.~331--348, 1984.

\bibitem{mckean}
H.~P. McKean, ``A class of {Markov} processes associated with nonlinear
  parabolic equations,'' {\em Proceedings of the National Academy of Sciences},
  vol.~56, pp.~1907--1911, Dec. 1966.

\bibitem{emrmodel}
E.~Pechersky, A.~Yambartsev, and V.~Zagrebnov, ``Stochastic dynamics of
  {Einstein} matter-radiation model with spikes,'' 2018.

\bibitem{longair}
M.~S. Longair, {\em High energy astrophysics}.
\newblock Cambridge university press, 2010.

\bibitem{macdon}
K.~B. Oldham, J.~C. Myland, and J.~Spanier, {\em The Macdonald Function Kv(x)},
  pp.~527--536.
\newblock New York, NY: Springer US, 2009.

\bibitem{lightman}
A.~P. {Lightman}, ``{Double {Compton} emission in radiation dominated thermal
  plasmas},'' {\em \apj}, vol.~244, pp.~392--405, Mar. 1981.

\bibitem{brown2}
L.~S. Brown, ``Compton scattering in a plasma,'' {\em Annals of Physics},
  vol.~200, no.~1, pp.~190--205, 1990.

\bibitem{peebles}
P.~J.~E. Peebles, L.~A. Page~Jr, and R.~B. Partridge, {\em Finding the {Big
  Bang}}.
\newblock Cambridge University Press, 2009.

\bibitem{silveira}
D.~d’Enterria and G.~G. da~Silveira, ``Observing light-by-light scattering at
  the large hadron collider,'' {\em Physical Review Letters}, vol.~111, no.~8,
  2013.

\bibitem{roy2}
D.~Roy, C.~M. Wilson, and O.~Firstenberg, ``Colloquium: Strongly interacting
  photons in one-dimensional continuum,'' {\em Rev. Mod. Phys.}, vol.~89,
  p.~021001, May 2017.

\bibitem{res}
G.~E.~F. Oliveira, C.~Maes, and K.~Meerts, ``Resetting photons,'' 2022.

\bibitem{gitlab}
K.~Meerts, ``Photon diffusion process.''
  \url{https://gitlab.kuleuven.be/u0131889/photon-diffusion-process/-/tags/v1.0}.

\end{thebibliography}

\end{document}